\documentclass[12pt]{article}
\usepackage{epsfig}
\usepackage{amssymb}
\def\ha{\mbox{$\frac{1}{2}$}}
\def\be{\begin{equation}}
\def\ee{\end{equation}}
\def\ba{\begin{array}{c}}
\def\ea{\end{array}}

\def\ben{\[}
\def\een{\]}

\newcommand{\bea}{\begin{eqnarray}}
\newcommand{\eea}{\end{eqnarray}}

\newcommand{\kt}{\rangle}
\newcommand{\br}{\langle}
\begin{document}

\begin{center}

{\Large \bf {

Mathematical and physical meaning of the crossings of
energy levels in ${\cal PT}-$symmetric systems

 }}


\vspace{20mm}

 {\bf Denis I. Borisov}

 \vspace{3mm}
Institute of Mathematics CS USC RAS, Chernyshevskii str., 112, Ufa,
Russia, 450008

and

Bashkir State Pedagogical University, October Rev. st., 3a, Ufa,
Russia, 450000

{e-mail: BorisovDI@yandex.ru }

\vspace{3mm}

and


\vspace{3mm}

 {\bf Miloslav Znojil}

 \vspace{3mm}
Nuclear Physics Institute ASCR, Hlavn\'{\i} 130, 250 68 \v{R}e\v{z},
Czech Republic

{e-mail: znojil@ujf.cas.cz}

\vspace{3mm}



\end{center}

\vspace{5mm}

\newpage
\section*{Abstract}


Unavoided crossings of the energy levels
due to a variation of a
real parameter
are studied. It is found that
after the quantum system
in question passes through one of its energy-crossing points
{\it alias} Kato's exceptional points (EP),
its physical
interpretation may {\em dramatically} change even when the
crossing energies themselves do not complexify. The anomalous
physical phase-transition mechanism of the change is revealed,
attributed to the EP-related mathematics and
illustrated via several
exactly solvable matrix toy models.

\section*{keywords}

quantum theory; real Kato's exceptional points; exactly solvable
models; anomalously broken ${\cal PT}-$symmetry; anomalous phase
transitions;

\newpage

\section{Introduction}

\subsection{Crossings of energies in systems with self-adjoint Hamiltonians}

One-parametric quantum Hamiltonians $\tilde{H}(\lambda)$ are most
often assumed self-adjoint inside a real interval of $\lambda \in
{\cal D}_{(physical)}$. This implies that an
unavoided crossing of energy levels is either excluded or
``incidental'', i.e., resulting from a symmetry. The centrally
symmetric harmonic oscillator with energies $\tilde{E}_{n,\ell}\sim
4n+2\ell+3$ where $n=0,1,\ldots$ and $\ell = 0, 1, \ldots$ may be
recalled as the best known illustration of the incidental
degeneracy due to which one has $\tilde{E}_{n+1,0}=\tilde{E}_{n,2}$,
etc.

\begin{figure}[h]                     
\begin{center}                         
\epsfig{file=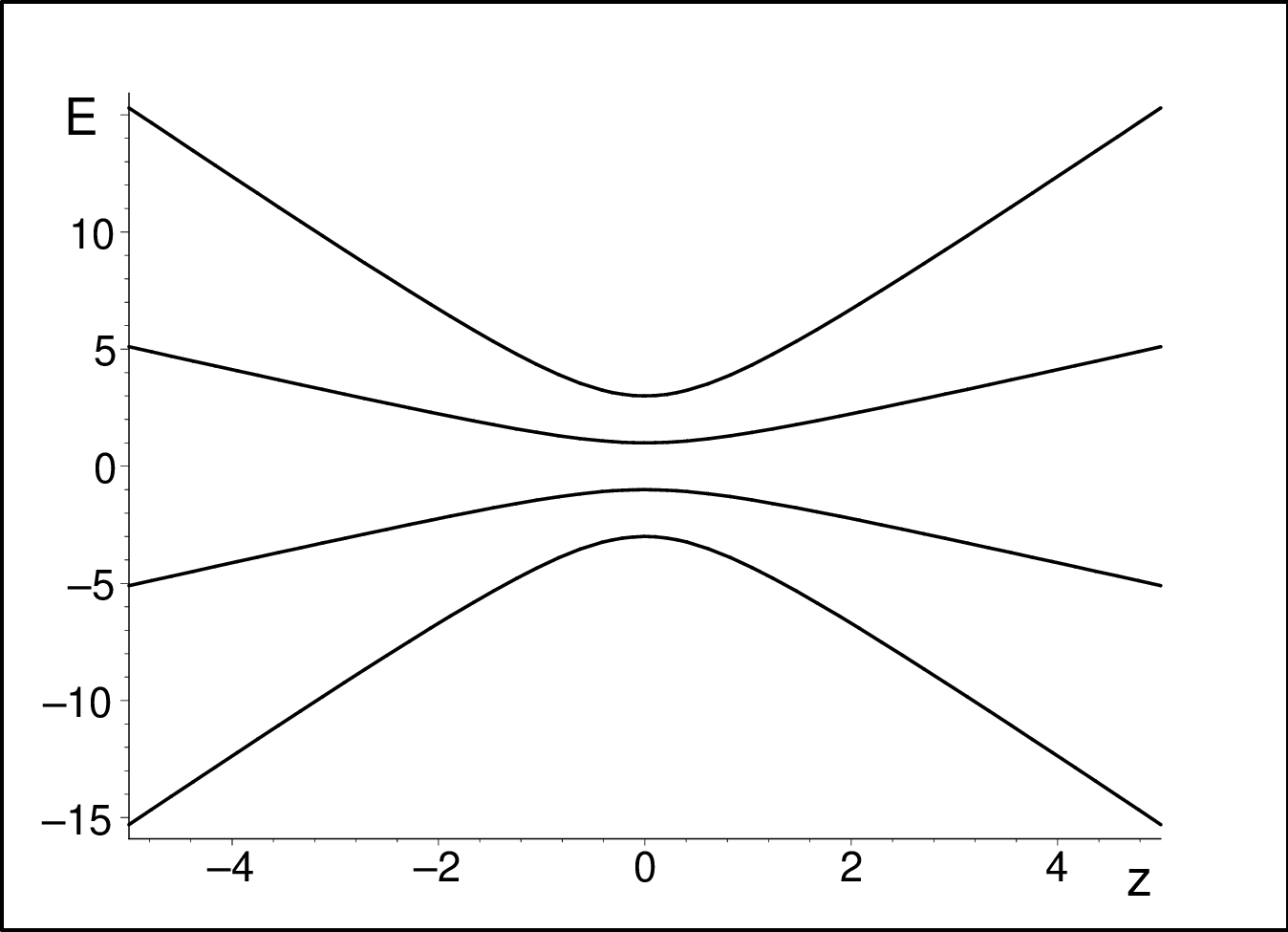,angle=270,width=0.55\textwidth}
\end{center}                         
\vspace{-2mm}\caption{Repulsion of levels for Hermitian Hamiltonian
(\ref{obilnice}).
 \label{obili}}
\end{figure}

The exclusion of
degeneracy accompanied by the well known tendency of eigenvalues to
avoid each other may be illustrated via the following four by four
tilded matrix
 \be
 \tilde{H}^{(4)}(z)=\left[ \begin {array}{cccc}
 -3&\sqrt{3}z&0&0\\
 \noalign{\medskip}\sqrt{3}z&-1&2\,z&0\\
 \noalign{\medskip}0&2\,z&1&\sqrt {3}z
 \\
 \noalign{\medskip}0&0&\sqrt {3}z&3
 \end {array} \right]
 =\left [
 \tilde{H}^{(4)}(z)
 \right ]^\dagger\,.
 \label{obilnice}
 \ee
This model without incidental symmetries nicely illustrates a
``mutual repulsion'' of eigenvalues (cf. Fig.~\ref{obili}).


\subsection{Crossings of energies  in ${\cal PT}-$symmetric models}

Incidental energy-level crossings also occur for multiple non-Hermitian
Hamiltonians exhibiting parity-times-time-reversal (a.k.a. ${\cal
PT}$, i.e., nonlinear) symmetry (cf. review paper \cite{Carl} or
recent papers \cite{Ahmed,Borisov}). One of the simplest
illustrations is provided by the generalized radial harmonic
oscillator Hamiltonian of Ref.~\cite{ptho}, i.e., by the
non-selfadjoint ordinary differential operator
 \be
 H^{(HO)}(\alpha,c)= -\,\frac{d^2}{dx^2} + x^2 -2ic\,x
 + \frac{\alpha^2-1/4}{(x-ic)^2}\,,
 \ \ \ \
 x \in (-\infty,\infty)\,
 \label{SEho}
 \ee
defined in $L^2(\mathbb{R})$ and possessing all of its energy
eigenvalues in closed form,
 \be
 E=E_{(n,q)}=4n+2 - 2 q \alpha+c^2\,,
 \ \ \ n=0,1,\ldots\,,
 \ \ \ q=\pm 1\,.
 \label{leverkussen}
 \ee
These quantities are real along the whole real line of
$\alpha$ (we may
ignore here the role of the inessential second parameter $c\neq 0$).
The unavoided energy-level crossings abound. At all of the integer
couplings $\alpha=m-n$ they have the form of degeneracies $E_{(m,1)}
=E_{(n,-1)}$.

\subsection{Exceptional points}

Tentatively, one could conjecture that in the context of crossing of
levels the linear and nonlinear symmetries might have played a
similar role. A deeper study of solvable models reveals that it is
not so. A number of decisive differences emerges. First of all,
Hermitian Hamiltonians exhibiting a linear symmetry remain
diagonalizable at the crossing point. In our non-Hermitian model
(\ref{SEho}), in contrast, all of the energy-degeneracy parameters
$\alpha=m-n$ are ``exceptional points'' (EP; the concept was
introduced by Kato \cite{Kato}) at which the Hamiltonian {\em
ceases} to be diagonalizable (see Ref.~\cite{ptho} for details). For
this reason the model {\em does not} admit the standard physical
probabilistic interpretation at any energy-crossing value of
$\alpha=m-n=\alpha^{(EP)}$. In contrast to their Hermitian
analogues, operators $H^{(HO)}(\alpha^{(EP)},c)$ {\em cannot}
consistently describe a quantum system. This means that the physics
which is controlled by a parameter may change abruptly at the EP
horizon \cite{catast}.

\begin{figure}[h]                     
\begin{center}                         
\epsfig{file=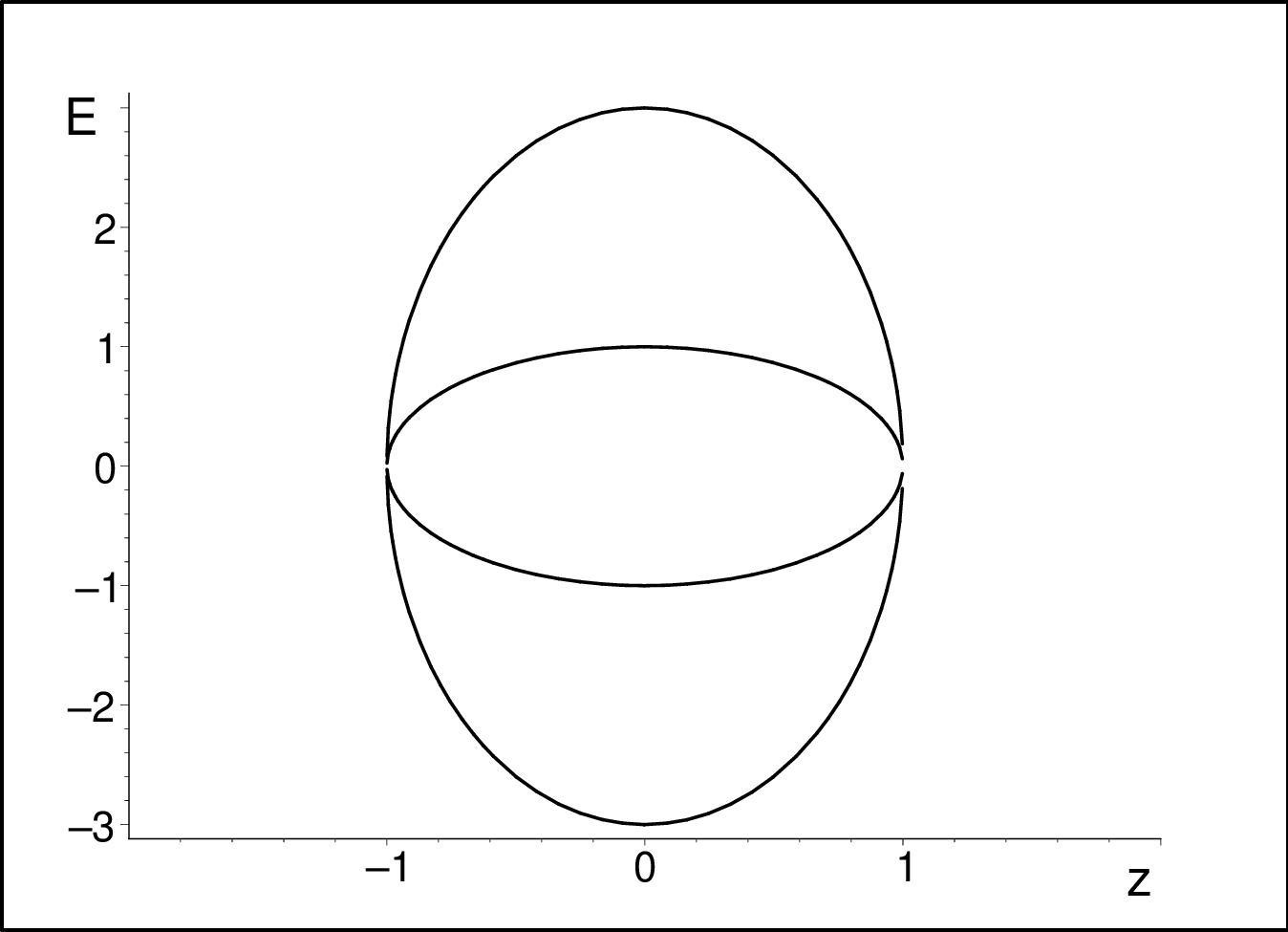,angle=270,width=0.55\textwidth}
\end{center}                         
\vspace{-2mm}\caption{Attraction of levels for non-Hermitian Hamiltonian
(\ref{jebilnice}).
 \label{elli}}
\end{figure}

The argument may further be strengthened when one recalls
the finite-dimensional and non-Hermitian ${\cal PT}-$symmetric
toy-models of Ref.~\cite{maximal}. Their four-by-four sample
 \be
 {H}^{(4)}(z)=\left[ \begin {array}{cccc}
  -3&\sqrt{3}z&0&0\\
 \noalign{\medskip}-\sqrt{3}z&-1&2\,z&0\\
 \noalign{\medskip}0&-2\,z&1&\sqrt {3}z
 \\
 \noalign{\medskip}0&0&-\sqrt {3}z&3
 \end {array} \right]
 \neq \left [
 {H}^{(4)}(z)
 \right ]^\dagger\,
 \label{jebilnice}
 \ee
differs from  (\ref{obilnice}) just by the
inversion of the signs in the lower diagonal. The new model is also
solvable yielding equidistant spectrum $E_n(z)=d_n\,\sqrt{1-z^2}$
with coefficients $d_0=-3$, $d_1=-1$, $d_2=1$ and $d_3=3$.
These energies are only real for $|z| \leq 1$ (cf. Fig.~\ref{elli}).
The two points $z_{coll.}=\pm 1$ of the collision of the
eigenvalues become exceptional in the sense of Kato, $z_{coll.}=
z^{(EP)}$. At these parameters the eigenvectors cease to form a complete basis.
This means that also mathematics changes abruptly at the EP horizon.

One of the most characteristic generic features  of
finite-dimensional non-Hermitian Hamiltonians exhibiting ${\cal PT}$
symmetry lies in an effective {\em attraction} between eigenvalues.
For  model (\ref{jebilnice}), in particular, the four half-hyperbolas
of Fig.~\ref{obili} become replaced by four half-ellipses of Fig.~\ref{elli}
(matched in
two ellipses). The whole spectrum is complex at all $z<-1$
and $z>1$.  {\it A priori}, no space seems left for a real
crossing of the levels. Other toy models must be sought.

\section{{\em Ad hoc}
physical Hilbert
spaces}

Our forthcoming considerations will be motivated by all of the
latter observations. We feel addressed by the
apparent lack of suitable (i.e., preferably, non-numerical) $N$ by
$N$ matrix examples which would exhibit an unavoided energy-level
crossing phenomenon (without complexifications) and which would
admit a consistent probabilistic
quantum-mechanical interpretation, i.e.,
an explicit construction of some standard
physical Hilbert space ${\cal H}^{(S)}$
of quantum states.
Our interest in
models with $N < \infty$ was also co-evoked by the
technical complexity
of the latter task in the case of $N=\infty$
\cite{117,Zelezny,unbali}.

\subsection{The concept of metric operator $\Theta$}


A given diagonalizable Hamiltonian with real spectrum may be found
non-Hermitian when considered in an unphysical Hilbert space ${\cal
H}^{(F)}$. In the notation of Ref.~\cite{SIGMA} the superscript
stands here for {\em both} ``false'' and ``favored'' {\it alias}
``friendly''. The most straightforward amendment of
the situation may be mediated by the replacement of the unphysical
Hilbert space by a physical one, ${\cal H}^{(F)} \to {\cal
H}^{(S)}$. This replacement is being realized by the
mere change of the inner product,
 \be
 \br \psi_1|\psi_2\kt^{(F)}\ \to \ \br \psi_1|\psi_2\kt^{(S)}
 =\br \psi_1|\Theta|\psi_2\kt^{(F)}\,
 \ee
where symbol $\Theta$ denotes the so called inner-product-metric
operator \cite{Geyer}.

The main idea of the recipe is that for a given Hamiltonian with
real spectrum which appeared non-Hermitian in ${\cal H}^{(F)}$ (we
will write $H \neq H^\dagger$) we may achieve, via a suitable choice
of metric, its Hermiticity in ${\cal H}^{(S)}$ (we will define
$H^\ddagger=\Theta^{-1} H^\dagger\Theta$ and write $H =H^\ddagger$).
The assignment of the Hermitizing
metric $\Theta$ to a given Hamiltonian $H$ is not unique \cite{Geyer}. This
ambiguity may play the role of a new freedom in quantum
model-building.

From an opposite perspective, a
unique choice of physical metric $\Theta$ enables us to decide
whether a given candidate for an observable is acceptable (i.e.,
Hermitian in given ${\cal H}^{(S)}$) or not.
Any change
of the metric would induce the change of the set of the operators of
observables, i.e., of the whole physical meaning and interpretation
of the quantum system in question. This idea will form a
background of our forthcoming considerations.

\subsection{Constructive specification of eligible metrics}

The concrete specification and practical use of metric $\Theta$ must
take into consideration its necessary mathematical properties
\cite{Geyer}. Firstly, in a setting valid for all observables, the
generator $H$ of the time evolution of wave functions must be
Hermitian in ${\cal H}^{(S)}$, i.e.,
 \be
 \sum_{k=1}^N\,
 \left [
      H^\dagger_{jk}\,\Theta_{kn}
      -\Theta_{jk}\,H_{kn}\right ] =0
 \,,\ \ \ \ \ j,n=1,2,\ldots,N\,,
 \ \ \ \ \ \ N={\rm dim}\ {\cal H}^{(F,S)} \leq \infty
   \,.
 \label{htot}
 \ee
Although  $H$ may be non-Hermitian in ${\cal H}^{(F)}$ (though not
necessarily -- see \cite{shendr}), the spectrum must be real in a
suitable physical domain ${\cal D}$ of a multiplet of parameters $\vec{\lambda}$.
Inside this domain, our preselected Hamiltonian $H=H(\vec{\lambda})$
must be also diagonalizable \cite{ali}. For the sake of
non-triviality of our considerations, we shall also assume the
non-emptiness of the EP boundary,  $\partial {\cal D}\neq
\emptyset$.

The spectrum of $H$ is often postulated non-degenerate,
discrete and bounded from below. This is a technical condition which
may easily be satisfied whenever one works with Hilbert spaces
${\cal H}^{(F)}$ of a finite dimension $ N < \infty$. In such a case
one may construct the (complete) set of $N$ eigenstates $|\Xi_j\kt$
of the F-space-conjugate operator $H^\dagger(\vec{\lambda})$,
 \be
 H^\dagger\,|\Xi_n\kt = E_n\,|\Xi_n\kt\,,\ \ \ \ n = 0, 1, \ldots, N-1\,.
 \label{sequa}
 \ee
Following Refs.~\cite{SIGMAdva}, we finally define the general
metric as the following sum
 \be
 \Theta=\Theta(H,\vec{\kappa})=\sum_{j=0}^{N-1}\,
 |\Xi_j\kt\,\kappa_n\,\br \Xi_j|\,.
 \label{metrr}
 \ee
The practically unrestricted variability of the optional parameters
$\kappa_j>0$ represents just the well known degree of
freedom of the theory.


\subsection{$N=2$ illustration}

In a two-by-two-matrix illustration using
real Hilbert space ${\cal H}^{(F)}=\mathbb{R}^2$, the
Hamiltonian-simulating matrix
 \be
 H=H^{(2)}(\lambda)=
 \left (
 \begin{array}{cc}
 0&1\\
 1+\lambda&0
 \ea
 \right )\,,\ \ \ \ \ \ \lambda>-1
 \label{wright}
 \ee
is exclusively Hermitian at $\lambda=0$ but it possesses
manifestly real and non-degenerate eigenvalues $E_\pm =\pm
\sqrt{1+\lambda}$ at any $\lambda>-1$. We may recall
Eq.~(\ref{metrr}) and define the general metric
 \be
 \Theta=\Theta^{(S)}(\lambda,b)=
 \left (
 \begin{array}{cc}
 1+\lambda&b\\
 b&1
 \ea
 \right )\,,\ \ \ \ -\sqrt{1+\lambda}<
 b <\sqrt{1+\lambda}
 \label{wrighter}
 \ee
with two positive eigenvalues $\theta_\pm
=1+\lambda/2\pm\sqrt{b^2+\lambda^2/4}$. This enables us to declare
{\em the same} Hamiltonian matrix (\ref{wright}) Hermitian in {\em
all\,}  Hilbert spaces ${\cal H}^{(S)}$
numbered by parameter $b$.

\section{Four-state non-Hermitian toy model}

Practical applications of
nontrivial metrics $\Theta$
suffer from a
scarcity of their supply \cite{MZbook}.
Up to rare exceptions \cite{David} a restriction of
attention to {\em finite} Hilbert-space dimensions $N < \infty$
seems necessary.
In a search for insight,
the use of the smallest $N$s  admitting non-numerical results
seems particularly rewarding.
Let us start, therefore, from the choice of $N=4$.

\subsection{Energies}

Illustrative Hamiltonian
(\ref{jebilnice}) was designed
as an example in which the spontaneous breakdown
of ${\cal PT}-$symmetry proceeds exclusively via
complexifications of the energies \cite{maximal}.
Such a model would be
unsuitable for our present purposes. Fortunately, in the light of
our more recent methodical studies \cite{Borisov,Wu} it appeared
that many methodical advantages of the family of $N$ by $N$ models
of Ref.~\cite{maximal} (like the reality of spectrum
or its non-numerical tractability) may be shared by simpler, albeit
more-parametric models in which the main diagonal is allowed to
vanish.
After we picked up the first nontrivial
two-parametric element
 \be
 H=H^{(4)}(\alpha,\beta)=   \left[ \begin {array}{cccc}
 0&-1+\beta&0&0\\\noalign{\medskip}-1-
 \beta&0&-1+\alpha&0\\\noalign{\medskip}0&-1-\alpha&0&-1+\beta
 \\\noalign{\medskip}0&0&-1-\beta&0\end {array} \right]\,
 \label{hamal}
 \ee
of this family (cf. Ref.~\cite{Wu}), we discovered that it may
offer the service.

The potentially
observable bound-state energies of model (\ref{hamal}) coincide with
the four real roots of secular equation
 \be
 {{\it E}}^{4}+ \left( {\alpha}^{2}-3+2\,{\beta}^{2} \right)
  {{\it E}}^{2}+1-2\,{\beta}^{2}+{\beta}^{4} =0\,.
  \label{secn4}
 \ee
These energies occur in pairs $E_{\pm,\varepsilon}=\pm
\sqrt{Z_\varepsilon}$ numbered by $\varepsilon=\pm$ where the symbol
$Z_\varepsilon$ denotes two easily written roots of a quadratic
equation. Inside the closure of the physical parametric
domain ${\cal D}$ these roots must be non-negative.

From the secular equation one immediately deduces the double
degeneracy $E \to 0$ of one of the pairs of the eigenenergies in the
limit of ${\beta}^{2}\to 1$. Under this constraint the complete
quadruple degeneracy $E_{\pm,\pm} \to 0$ takes place in the second
limit of ${\alpha}^{2}\to 1$. Still, the exact knowledge of the
energies
 $$
 E_{\pm,\pm}=\pm \ha\,\sqrt {6 -2\,{\alpha}^{2}-4\,{\beta}^{2}
 \pm 2\,\sqrt{{\alpha}^{4}-6\,{\alpha}^{2}
 +4\,{\alpha}^{2}{\beta}^{2}+5-4\,{\beta}^{2}}}\,
 $$
offers more insight than expected.

\subsection{A reparametrization }

In terms of new variables $A=1-\alpha^2$, $B=1-\beta^2$ and $C=A+4B$
the previous formula becomes more transparent,
 \be
 2\,E_{\pm,\pm}=\pm \sqrt {A+C \pm 2\,
  \sqrt{AC}}=\pm \sqrt {(\sqrt{A} \pm \sqrt{C})^2}
  = \pm \sqrt{A} \pm\sqrt{C}\,.
   \label{levecomp}
 \ee
The reparametrization clarifies
the root-complexification nature of the lines $A=0$ and $C=0$.
More precisely, formula~(\ref{levecomp}) indicates that the set of
the potentially physical parameters $A$ and $B$ or $C$ yielding the
real spectrum of energies is specified by the two elementary
inequalities $A\geq 0$ and $C\geq 0$ in the $A-B$ plane (cf.
Fig.~\ref{fia1}).

\begin{figure}[h]                     
\begin{center}                         
\epsfig{file=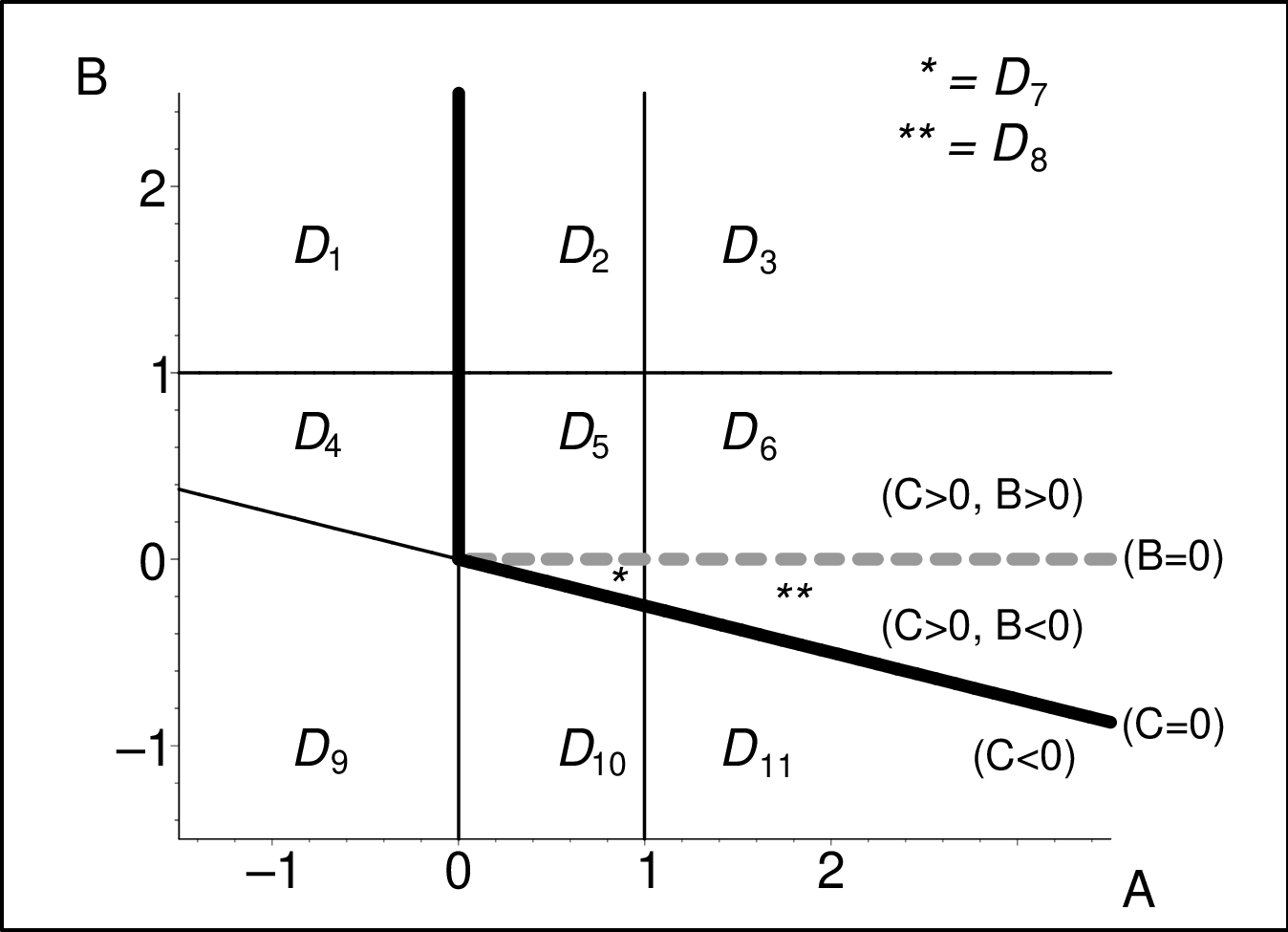,angle=270,width=0.55\textwidth}
\end{center}                         
\vspace{-2mm}\caption{The $A-B$ plane of reparametrized Hamiltonian
(\ref{hamal}). After exemption of the dashed line, the points inside
the thick-line-bounded upper-right wedge specify the unitary
dynamical regime.
 \label{fia1}}
\end{figure}


After a return to the old parameters $\alpha= \sqrt{1-A}$ and
$\beta= \sqrt{1-B}$, our new $N=4$ matrix (\ref{hamal}) would cease
to be real in the whole $A-B$ plane. This slightly redefines the model.
Keeping this
in mind let us further recall Fig.~\ref{fia1} and separate the $A-B$
plane of parameters into eleven subdomains while noticing that


\begin{itemize}

\item
in the usual matrix sense, i.e., inside the most common
complex vector space ${\cal H}^{(F)}\equiv \mathbb{C}^4$ endowed
with trivial metric $\Theta^{(F)}=I$, our (possibly, complex)
Hamiltonian (\ref{hamal}) is manifestly Hermitian just in the
single subdomain
${\cal D}_3$;

\item
our four by four Hamiltonian is a real
matrix with real spectrum just in the two simply connected
subdomains of parameters ${\cal D}_5$ and ${\cal D}_7$;

\item
the spectrum is real inside the
closure of the union
 $\overline{{\cal D}_2\cup{\cal D}_3\cup{\cal D}_5\cup{\cal D}_6
 \cup{\cal D}_7\cup{\cal D}_8}\,
 $ of six subdomains.

\end{itemize}

\begin{figure}[h]                     
\begin{center}                         
\epsfig{file=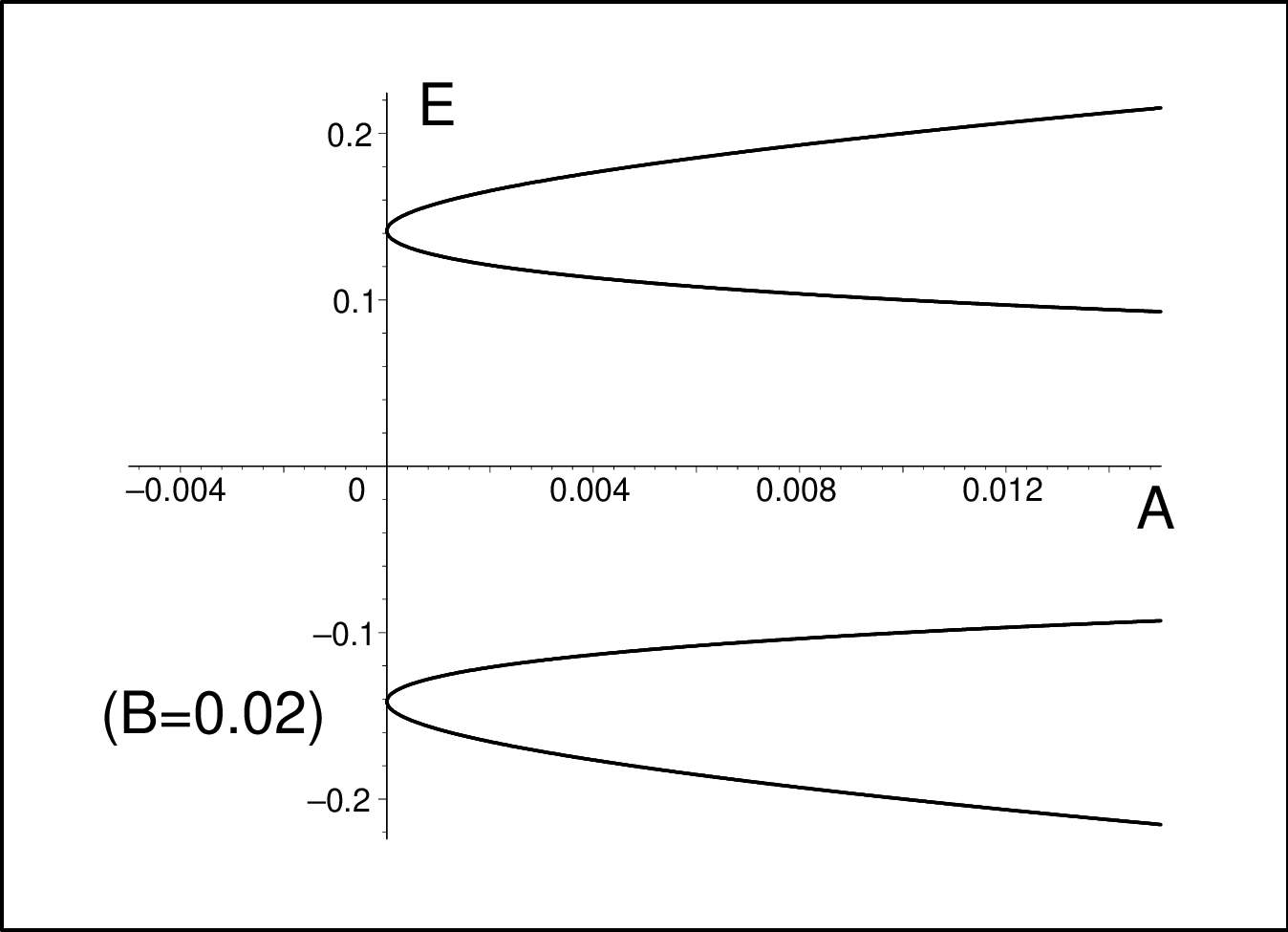,angle=270,width=0.55\textwidth}
\end{center}                         
\vspace{-2mm}\caption{The spectrum in the vicinity of the $A=0$
horizon of the first kind.
 \label{fia1v}}
\end{figure}


 \noindent
In Fig.~\ref{fia1} the two thick EP half-lines with  $A=0$ and
$C\geq 0$ or with $C=0$ and $A \geq 0$ play the role of the
boundaries of stability of the system (let us call them ``quantum
horizons of the first kind''). Beyond these horizons the energies
complexify and cease to be observable.

The most elementary illustration of this most common form of quantum
phase transition is provided by Fig.~\ref{fia1v} where we varied
parameter $A$ along a line connecting the unphysical subdomain
${\cal D}_4$ with  its most conventional physical neighbor ${\cal
D}_5$. Once we choose a nonvanishing second parameter $B=1/50$ we
obtained a generic picture in which the two separate degenerate
energies are unfolding in parallel.

With the decrease of $B>0$ the degenerate energies get closer to
each other. In the limit one arrives at an exceptional,
double-degeneracy scenario with $A=B=0$. The spectrum in the
vicinity is sampled in Fig.~\ref{fia1u}. One moves there
along the path with $B=A$ so that the system passes through the
origin in a way connecting the physical region ${\cal D}_5$ with the
twice-forbidden unphysical subdomain ${\cal D}_9$. Obviously, one
could now reinterpret a return to the pattern of Fig.~\ref{fia1v} as
a consequence of perturbation due to which the upper and lower
doublets get decoupled.

\begin{figure}[h]                     
\begin{center}                         
\epsfig{file=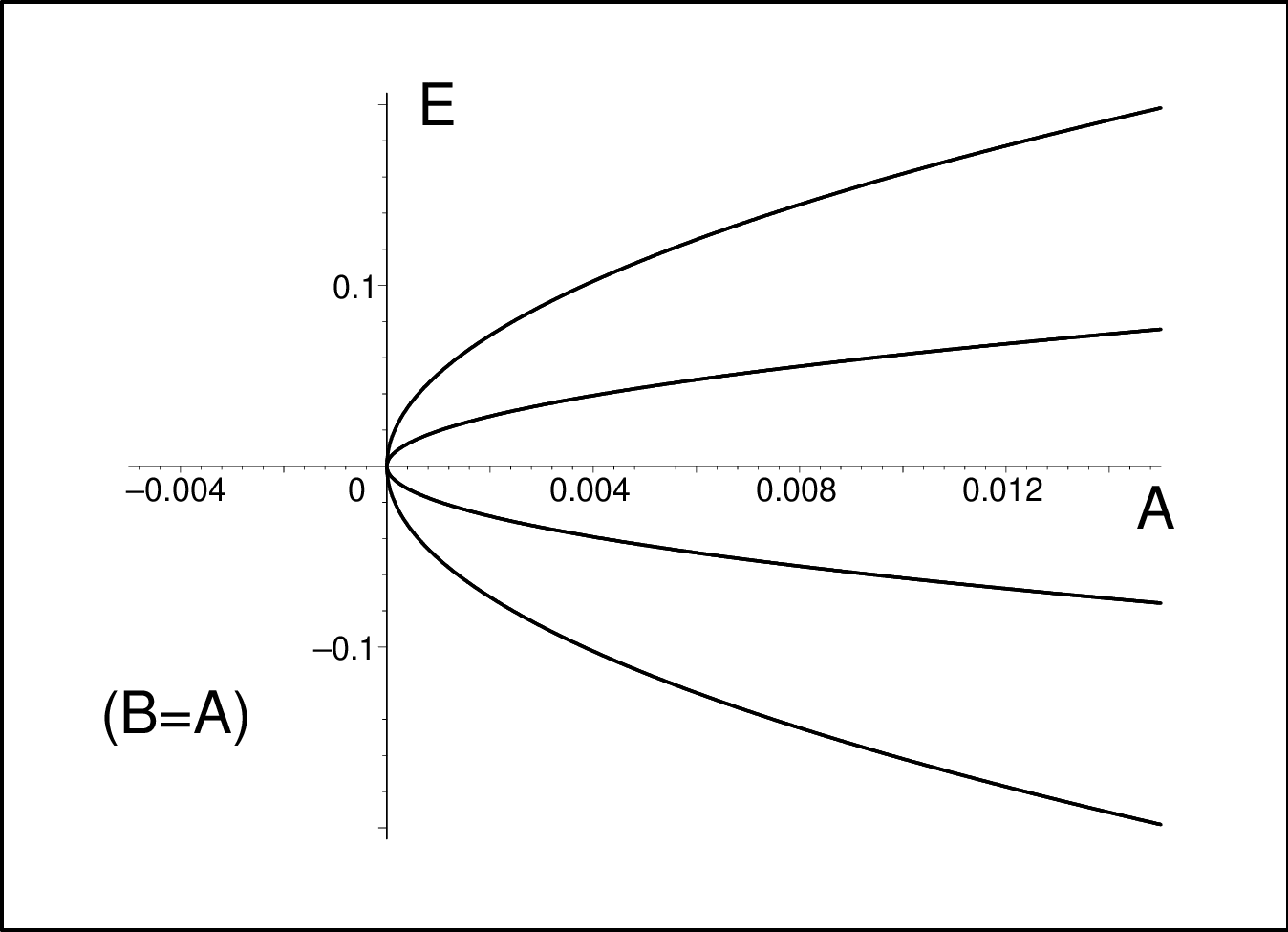,angle=270,width=0.55\textwidth}
\end{center}                         
\vspace{-2mm}\caption{The confluent-EP scenario at $B=A$.
 \label{fia1u}}
\end{figure}


The $B=A$  pass is anomalous because
inside the twice-forbidden subdomain ${\cal D}_9$ the model happens
to have a purely imaginary spectrum. As long as this means
that ${\rm Im}\ ({\rm i}\,E_n)=0$, one could obtain a potentially
measurable spectrum also in subdomain ${\cal D}_9$, using simply a
premultiplied form $\hat{H}={\rm i}\,H^{(4)}(\alpha,\beta)$ of an
after-transition candidate for one of possible physical Hamiltonians
in ${\cal D}_9$.

\section{New physics behind the unavoided level crossings}


Admitting, in Fig.~\ref{fia1}, a further decrease of $B$ below zero
while keeping $A \geq 0$ we enter another dynamical regime which
opens the possibility of the $C=0$ EP phase transitions of the first
kind. During them one moves, typically, from the physical subdomain
${\cal D}_7$ to its unphysical neighbor ${\cal D}_{10}$. The
parameter-dependence of the spectrum as well as its complexification
pattern will be analogous to the ones displayed in
Fig.~\ref{fia1v}.

Along both of the thick EP
lines of Fig.~\ref{fia1} the phase transitions between the complex
and real spectrum are qualitatively the same (i.e., in our
terminology, of the first kind). In both of these cases the
degeneracy of a pair of energies
at the EP singularity is followed by its subsequent unfolding
into unobservable complex eigenvalues. This mechanism is widely
known as the so called spontaneous breakdown
of ${\cal PT}$ symmetry (see also its
numerous exactly solvable models in \cite{Levai}).

What remains unclarified is the physical nature of the other, alternative
parameter-changing processes during which
a pair of energies would pass through the
remaining, dashed $B=0$ EP line of Fig.~\ref{fia1}
{\em without} getting complexified.
We intend to show now
that after one crosses such an EP horizon
there will emerge good reasons for speaking about
an anomalous phase transition
``of the second kind''.


\subsection{The menu of metrics}

In the light of formula (\ref{metrr}) the metric ceases to be
positive definite at any EP parameter. Keeping in mind
Fig.~\ref{fia1} we may conclude that no positive definite metric
$\Theta$ can exist at $A\leq 0$, at $C\leq 0$ and at $B= 0$.
Temporarily, let us assume that $A>0$, $C>0$ and $B \neq 0$,
therefore.

Once we insert Hamiltonian (\ref{hamal}) in the implicit linear
algebraic definition (\ref{htot}) of the real, symmetric and
positive definite metric matrix $\Theta$, we obtain an
overdetermined set of 16 equations for 10 unknown matrix elements.
As long as formula (\ref{metrr}) indicates that there are strictly
four free real parameters in the family of solutions, let us pick
up the quadruplet of elements $\Theta_{1j}=t_j$ with
$j=1,2,3,4$ as free parameters. Next, let
us solve the system by the standard elimination technique yielding
 $$
\Theta_{22}=-{\frac {-{\it {t}_1}+{\it {t}_1}\,\beta-{\it
{t}_3}-{\it {t}_3}\,\alpha}{1+\beta}}\,,
 \ \ \
\Theta_{23}={\frac {{\it {t}_2}-{\it {t}_2}\,\alpha+{\it {t}_4}+{\it
{t}_4}\,\beta}{1+\beta}}\,,
 \ \ \
\Theta_{24}=-{\frac {{\it {t}_3}\, \left( -1+\beta \right)
}{1+\beta}}\,
 $$
in the second row of the matrix,
 $$
\Theta_{33}={\frac {{\it {t}_1}-{\it {t}_1}\,\alpha-{\it
{t}_1}\,\beta+{\it {t}_1}\,\beta\, \alpha+{\it {t}_3}-{\it
{t}_3}\,{\alpha}^{2}}{1+\beta+\alpha+\alpha\,\beta}}=
 {\frac { \left( -{\it {t}_1}+{\it {t}_1}\,\beta-{\it {t}_3}-{\it
{t}_3}\,\alpha
 \right)  \left( -1+\alpha \right) }{ \left( 1+\beta \right)  \left( 1
+\alpha \right) }}\,,
 $$
 $$
\Theta_{34}={\frac {{\it {t}_2}\, \left(
1-\alpha-\beta+\alpha\,\beta \right) }{1+
\beta+\alpha+\alpha\,\beta}}=
 {\frac {{\it {t}_2}\, \left( -1+\alpha \right)  \left( -1+\beta
\right) } { \left( 1+\beta \right)  \left( 1+\alpha \right) }}
 $$
in the third row and
 $$
\Theta_{44}=-{\frac {{\it {t}_1}\, \left(
\alpha\,{\beta}^{2}-{\beta}^{2}+2\,\beta-2 \,\alpha\,\beta+\alpha-1
\right) }{{\beta}^{2}+\alpha\,{\beta}^{2}+2\,
\beta+2\,\alpha\,\beta+1+\alpha}}=
 -{\frac {{\it {t}_1}\, \left( -1+\beta \right) ^{2} \left(
 -1+\alpha
 \right) }{ \left( 1+\beta \right) ^{2} \left( 1+\alpha \right) }}
 $$
in the fourth row of the metric. An exhaustive,
general and complete solution is obtained. It would be too
space-consuming to display the whole matrix of the eligible metrics
in print. Still, its display element by element
enables us to discuss some of the most important consequences.

\subsection{EP horizon of the second kind}

The insertion of $B=0$ {\em alias} $\beta=1$ reduces
our Hamiltonian (\ref{hamal}) to  one-parametric matrix
 \be
 H= H^{(4)}(\alpha,1)= \left[ \begin {array}{cccc}
 0&0&0&0\\\noalign{\medskip}-2&0&-1+\alpha&0
 \\\noalign{\medskip}0&-1-\alpha&0&0
 \\\noalign{\medskip}0&0&-2&0\end {array} \right]\,.
 \label{hamalv}
 \ee
One can easily prove that such a matrix possesses two vanishing
eigenvalues $E=0$ but just a single related eigenvector. This means
that matrix (\ref{hamalv}) is non-diagonalizable and that the $B=0$
line is all composed of exceptional points.
The Jordan-block
canonical structure of the $B=0$ Hamiltonian cannot be Hermitized by
any metric $\Theta$. Two of the eigenvectors $|\Xi_j\kt$ in formula
(\ref{metrr}) coincide in the limit $B \to 0$ so that in the same
limit, one of the eigenvalues of  $\Theta$ goes to zero. All
of the metric-candidates of the concrete form (\ref{metrr}) become
non-invertible at $B=0$.

\begin{figure}[h]                     
\begin{center}                         
\epsfig{file=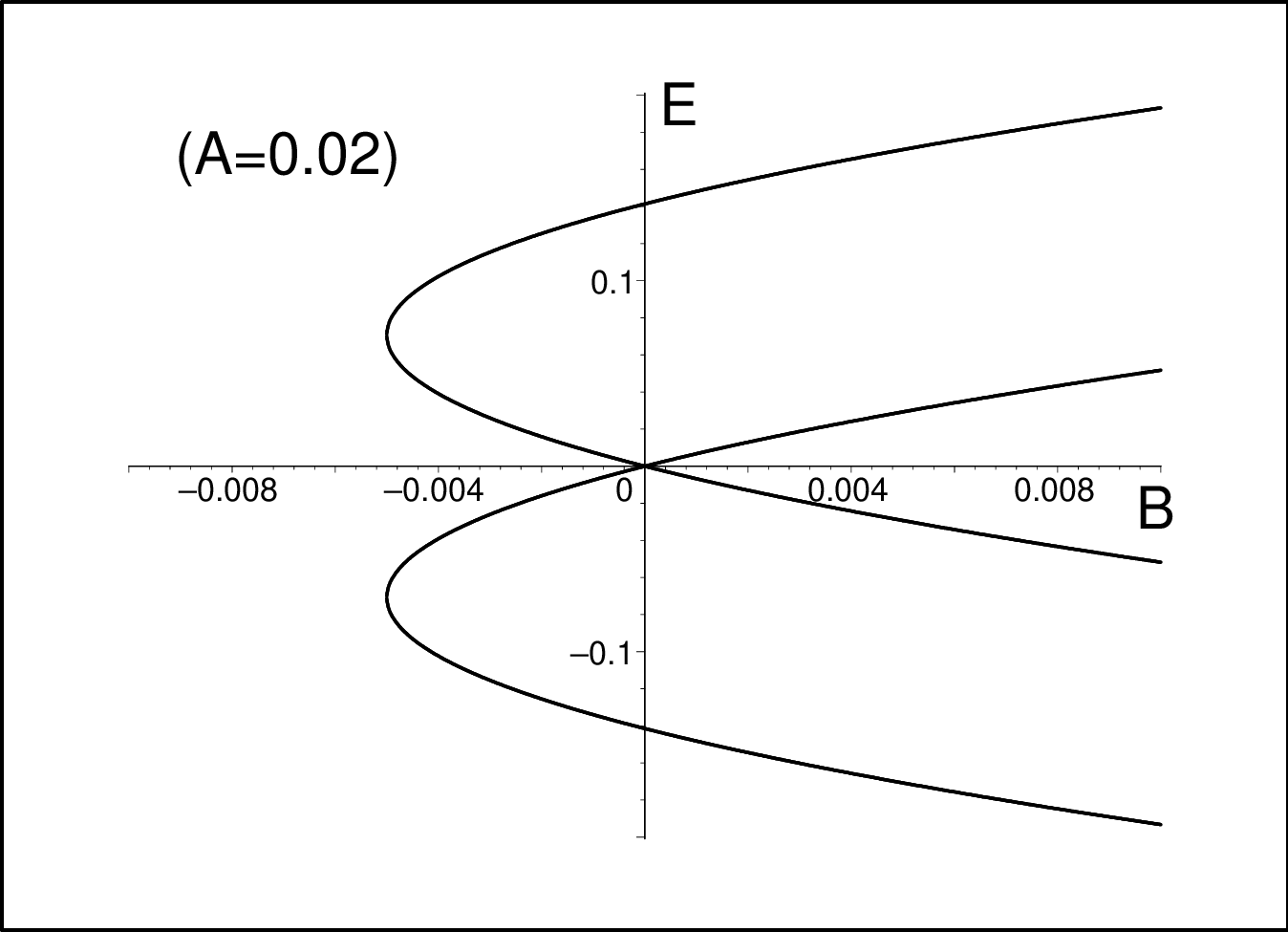,angle=270,width=0.55\textwidth}
\end{center}                         
\vspace{-2mm}\caption{The unavoided level crossing at $B=0$ for
$A=1/50$.
 \label{fia1w}}
\end{figure}


The
$B-$dependence of the energy levels is such that two of them merge
at $B = 0$. In the vicinity of the $B=0$ singularity (i.e., in our
present terminology, along the EP horizon of the second kind) one
observes the unavoided level crossing, the concrete form of which is
illustrated in Fig.~\ref{fia1w}.
The picture may be complemented by the closed-form construction of
the bound-state solutions starting from the small-perturbation
version
 $$
 H= H^{(4)}(\alpha,1-\gamma)
 =\left[ \begin {array}{cccc} 0&-\gamma&0&0\\\noalign{\medskip}-2+
 \gamma&0&-1+\alpha&0\\\noalign{\medskip}0&-1-\alpha&0&-\gamma
 \\\noalign{\medskip}0&0&-2+\gamma&0\end {array} \right]\,
 $$
of the original Hamiltonian. A small shift $\gamma$ in
$\beta=1-\gamma$ yields an equally small value of $B=2\gamma+{\cal
O}(\gamma^2)$ of both signs. The resulting closed form of the pair
of the almost-vanishing eigenvalues reads
 \be
 \pm 2 E_{\pm,-}
 =\sqrt {2-2\,{\alpha}^{2}+8\,\gamma-4\,{\gamma}^{2}-2\,\sqrt {{
\alpha}^{4}-8\,{\alpha}^{2}\gamma+4\,{\alpha}^{2}{\gamma}^{2}-2\,{
\alpha}^{2}-4\,{\gamma}^{2}+8\,\gamma+1}}\,.
 \ee
One quickly arrives at the required perturbation-expansion
description of the crossing phenomenon in the language of Taylor
series
 $$
 E_{\pm,-} \approx \pm  \left( 2+{\alpha}^
{2}+3/4\,{\alpha}^{4} +\ldots \right) \gamma \mp \left(
5+13/2\,{\alpha}^{2}+\ldots \right) {\gamma}^{2} \pm \ldots\,.
 $$
The change of the sign of the auxiliary small parameter $\gamma$ may
be perceived as a transition between the potentially physical
real-spectrum domain ${\cal D}_5$ and another, {equally acceptable}
real-spectrum domain ${\cal D}_7$.

\subsection{Phase transition of the second kind}

During the above-mentioned transition, the only suspicious point is
$B=0$ at which the metric ceases to exist. Hence, we have to analyze
the $B-$dependence of the metric near the EP singularity at $B=0$ in
a more explicit representation. Most efficiently, such a task may be
simplified when we accept the specific choice of $t_2=t_3=t_4=0$.
Under a symmetrized overall normalization choice of $t_1 \neq 0$
this makes our metric strictly diagonal, with elements
 \ben
 \Theta_{11}={\frac { \left( 1+\alpha \right)  \left( 1+\beta \right)
 }{1-\beta}}\,, \ \ \ \ \Theta_{22}=1+\alpha\,, \ \ \ \
 \Theta_{33}=1-\alpha\,, \ \ \ \ \Theta_{44}={\frac { \left(
 1-\alpha \right)  \left( 1-\beta \right) }{1+\beta}}\,.
 \een
%
Inside the physical subdomain
${\cal D}_5$ of Fig.~\ref{fia1} our diagonal metric is positive
definite for all of the real parameters such that $|\alpha|<1$ and
$|\beta|<1$. Below the EP line $B=0$ our metric
ceases to be positive definite.

As long as we stay
inside the physical domain giving real energies (viz., inside
subdomain ${\cal D}_7$ of Fig.~\ref{fia1}) we may put
$\beta=1+\delta^2$ (where $\delta$ is small but real) and check the
statement. It gets verified: our diagonal matrix $\Theta$ loses the
status of metric and becomes converted into the mere indefinite
diagonal pseudometric ${\cal P}$ which possesses two negative
elements and/or eigenvalues,
 \ben
 {\cal P}_{11}=-{\frac { \left( 1+\alpha \right)  \left( 2+\delta^2 \right)
 }{\delta^2}}\,, \ \ \ \ {\cal P}_{22}=1+\alpha\,, \ \ \ \
 {\cal P}_{33}=1-\alpha\,, \ \ \ \ {\cal P}_{44}=-\delta^2\,{\frac {
 1-\alpha   }{2+\delta^2}}\,.
 \een
Below the EP line $B=0$, {\em any} correct
physical metric must necessarily be non-diagonal.
The {\em physics} of the quantum system in question will be
{\em different} in the neighboring physical subdomains ${\cal D}_5$
and ${\cal D}_7$. The energies remain observable but
the set of the admissible operators of observables for parameters
inside ${\cal D}_5$ will necessarily be different from the set of
the operators of observables for parameters which crossed the $B=0$
line and belong to ${\cal D}_7$.

Such a change of physics at $B=0$ is not as drastic as
the truly catastrophic loss of the reality of the energies at the
horizons $A=0$ or $C=0$. Still, one must speak about phase
transition. We propose to call such a change the phase transition of
the second kind.

\section{Level crossings beyond $N = 4$\label{setri}}

When addressing conceptual matters we made an ample use, up to now,
of the elementary nature of the toy-model secular Eq.~(\ref{secn4})
at $N=4$. At a few higher matrix dimensions $N$
the determination of the EP horizons is more complicated
but still non-numerical. The
methods were described in Ref.~\cite{horizon} where, for a not too
dissimilar class of matrix models, these methods were shown
effective up to $N=11$.

\subsection{The family of models\label{ytsetri}}


The pass of a quantum observable (typically, of Hamiltonian
$H(\lambda)$) through a Kato's exceptional point $\lambda^{(EP)}$
leads, typically, to a quantum catastrophe during which certain
eigenvalues collide and, subsequently, complexify. The observability
status of Hamiltonian $H(\lambda)$ is lost and the critical value of
$\lambda=\lambda^{(EP)}$ may be perceived as a point on horizon of
quantum stability. In the alternative, eigenvalue-crossing scenario
without complexification we reminded the readers that one has to
distinguish between the non-EP degeneracy (typical for Hermitian
models) and an anomalous, EP-caused degeneracy.
In this general theoretical setting \cite{Borisov} we revealed that
one may encounter a loss of the system's
observability implying a subtler form of the quantum phase
transition.

Via the solvable $N=4$  example we discovered that the
mechanism of the anomalous transition is based on the loss of the
positivity of the metric at the EP singularity. The Hilbert space
(i.e., its inner product, i.e., the set of
the eligible operators of observales) changed. Beyond
the eigenvalue-collision at $\lambda=\lambda^{(EP)}$ the physical
contents of the theory may be entirely different even if the energy
spectrum itself stays real.

Whenever the matrix dimensions
get too large, the proofs become more and more numerical even
when we keep working with the most elementary tridiagonal and finite-dimensional
quasi-real matrix Hamiltonians of Ref.~\cite{Wu},
%
%
 \be
 H^{(N)}({\lambda, \mu,\ldots})=
 \left( \begin {array}{ccccccc}
  2&-1+{\it {\lambda}}&0&\ldots&&\ldots&0
 \\{}-1-{\it {\lambda}}&2&-1+{\it {\mu}}&0&\ldots&&\vdots
 \\{}0&-1-{\it {\mu}}&2&-1+{\nu}&0&\ldots&
 \\{}\vdots&0&-1-{\nu}&2&\ddots&\ddots&\vdots
 \\{}&&\ddots&\ddots&\ddots&-1+\mu&0
  \\{}\vdots&&&\ddots&-1-\mu&2&-1+{\it {\lambda}}
 \\{}0&\ldots&\ldots&0&0&-1-{\it {\lambda}}&2
 \end {array}
 \right)\,.
 \label{hlham}
 \ee
The kinematics may be perceived as represented by
the discrete Laplacean $T=H^{(N)}({0, 0,\ldots})$. The
information about the dynamics is carried by the set of
$N/2$ couplings.

Our preliminary numerical experiments with the $N>4$ models of the
above class proved encouraging, providing a few new qualitative
insights (cf. the next subsection). On the abstract level it was
useful that the interaction $V=H-T$ itself was kept minimally
non-local and antisymmetric. The choice was further restricted to
the matrices which were required  ${\cal PT}-$symmetric with respect
to the most elementary antidiagonal $N$ by $N$ parity-simulating
matrix
 \be
 {\cal P}= {\cal P}^{(N)}=
  \left[ \begin {array}{ccccc}
 0&0&\ldots&0&1
 \\{}0&\ldots&0&1&0\\
 {}\vdots&
 {\large \bf _. } \cdot {\large \bf ^{^.}}&
 {\large \bf _. } \cdot {\large \bf ^{^.}}
 &
 {\large \bf _. } \cdot {\large \bf ^{^.}}&\vdots
 \\{}0&1&0&\ldots&0
 \\{}1&0&\ldots&0&0
 \end {array} \right]\,
 \label{anago}
 \ee
in combination with the time-reversal-simulating antilinear operator
${\cal T}$ of matrix transposition.

\begin{figure}[h]                     
\begin{center}                         
\epsfig{file=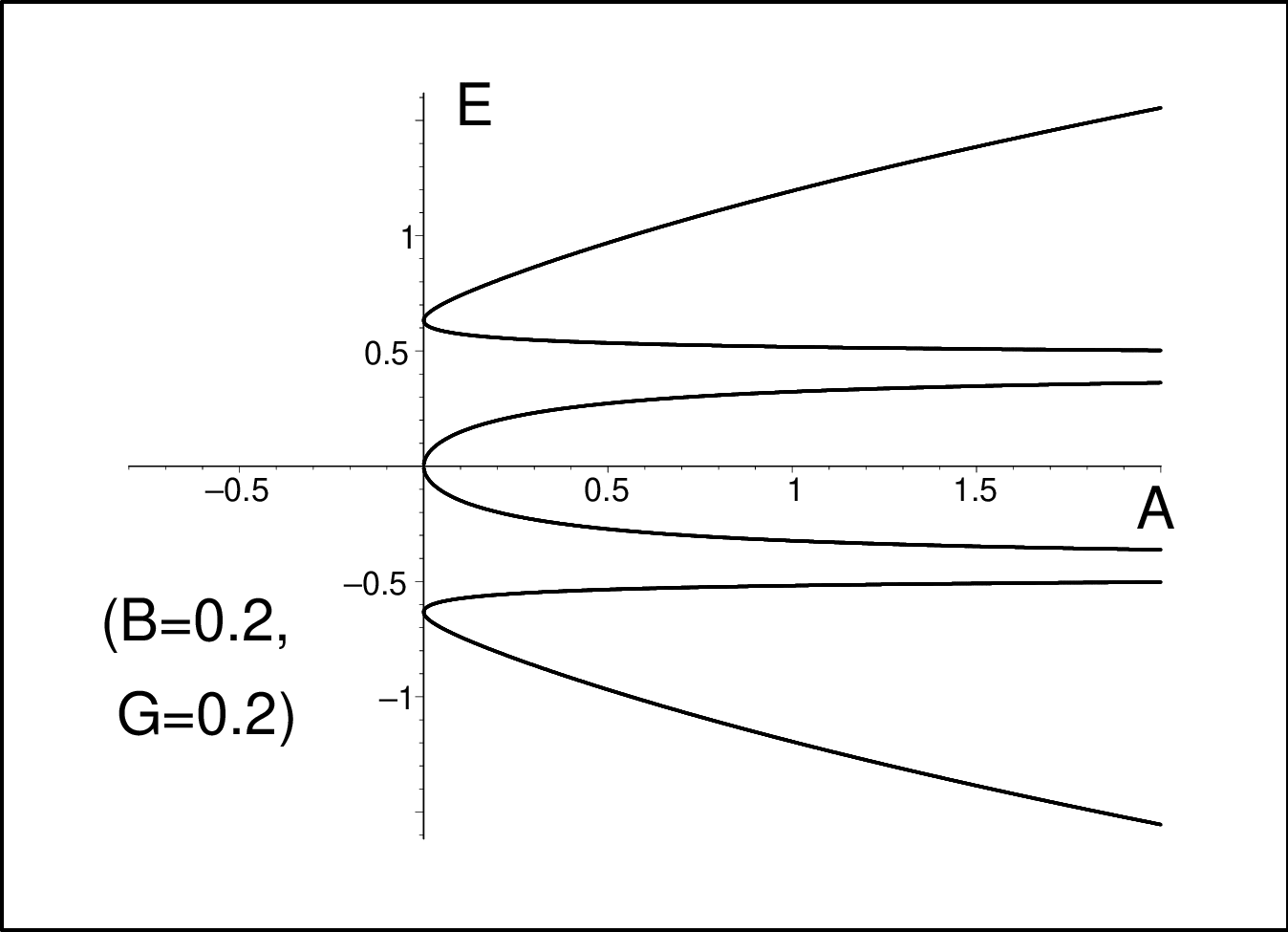,angle=270,width=0.55\textwidth}
\end{center}                         
\vspace{-2mm}\caption{The $N=6$ spectrum near the $A=0$ horizon of
the first kind.
 \label{fia1vb}}
\end{figure}
%

\subsection{Non-Hermitian quantum lattice with $N=6$}

The study of the three-parametric $N=6$ model
 \be
 H=H^{(6)}(\alpha,\beta.\gamma)=
 \left[ \begin {array}{cccccc} 0&-1+\gamma&0&0&0&0
\\\noalign{\medskip}-1-\gamma&0&-1+\beta&0&0&0
\\\noalign{\medskip}0&-1-\beta&0&-1+\alpha&0&0
\\\noalign{\medskip}0&0&-1-\alpha&0&-1+\beta&0
\\\noalign{\medskip}0&0&0&-1-\beta&0&-1+\gamma
\\\noalign{\medskip}0&0&0&0&-1-\gamma&0\end {array} \right]
 \label{heajmal}
 \ee
provides an insight
into the pattern of possible generalizations. Reparametrizations
$A=1-\alpha^2$, $G=1-\beta^2$ and $B=1-\gamma^2$
enable us to establish a connection between the $N=4$ and $N=6$ spectra.

\begin{figure}[h]                     
\begin{center}                         
\epsfig{file=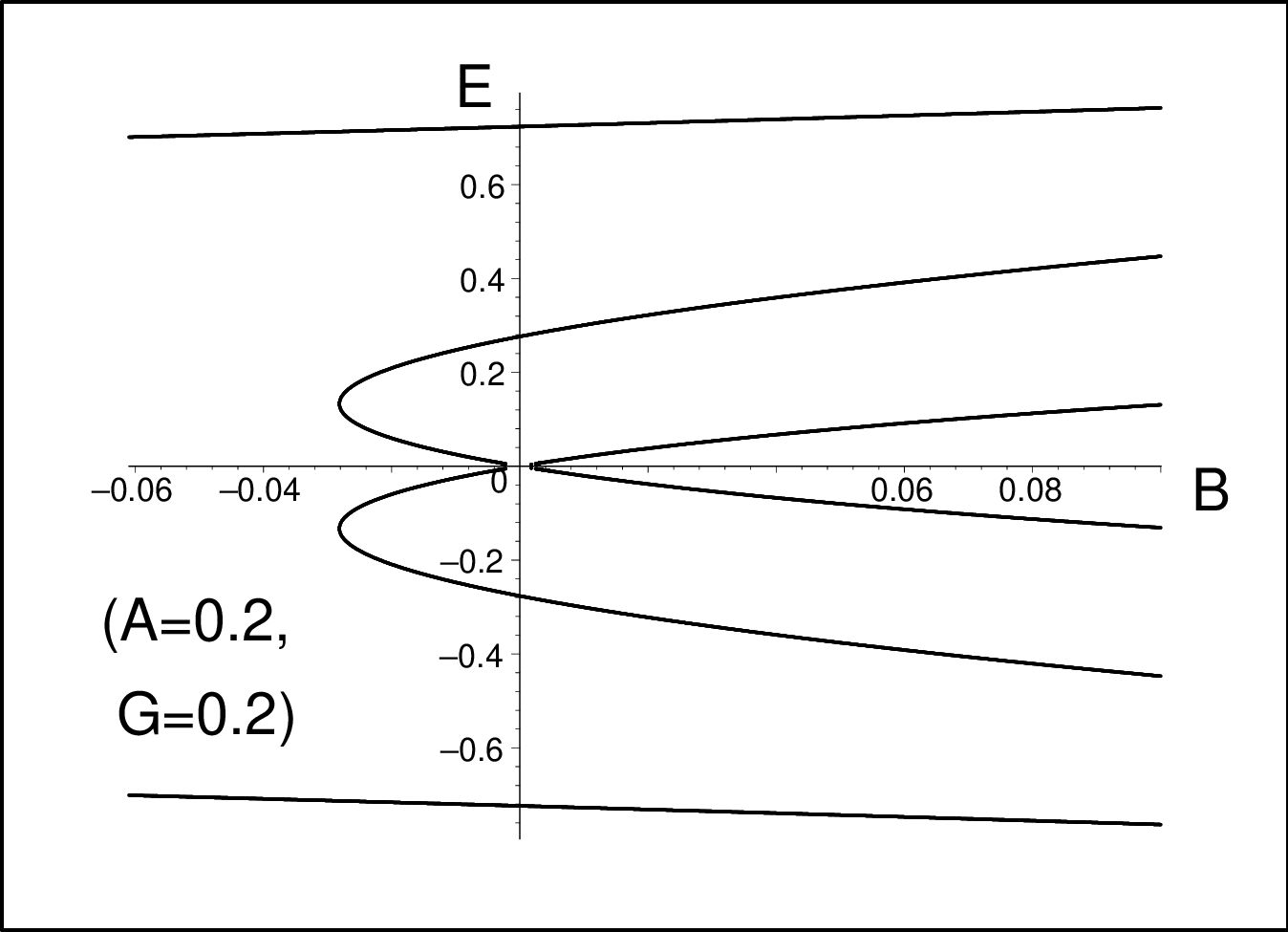,angle=270,width=0.55\textwidth}
\end{center}                         
\vspace{-2mm}\caption{The unavoided inner-level crossing at $B=0$
for $N=6$.
 \label{fia1wb}}
\end{figure}
%

\begin{itemize}

\item
 in the ``innermost coupling''
 dynamical regime
  we find the same no-intersection pattern both in
 Fig.~\ref{fia1v} (where $N=4$) and in Fig.~\ref{fia1vb} (where $N=6$);
 the same form of the phase transition of the first kind
 may be expected to survive at all of the higher dimensions $N<\infty$;

\item
 in the opposite, ``outermost coupling''
 dynamical regime
 the inner-level-crossing
  pattern
  (which characterizes the phase transition of the second kind)
  emerges both in
 Fig.~\ref{fia1w} (with $N=4$) and in
  Fig.~\ref{fia1wb} (with $N=6$);
 a very similar pattern may be expected at all  $N>6$;

\item
 in the newly emerging ``intermediate-coupling''
 dynamical regime
 the phase transition of the first kind is expected;
 in the first nontrivial $N=6$ example of Fig.~\ref{fia1vc}
 the $G=0$ EP mergers only
 involve  two pairs of levels while the
 reality of the remaining spectrum
 is not destroyed.
 This or similar pattern is also expected to occur at $N > 6$.

\end{itemize}

\begin{figure}[h]                     
\begin{center}                         
\epsfig{file=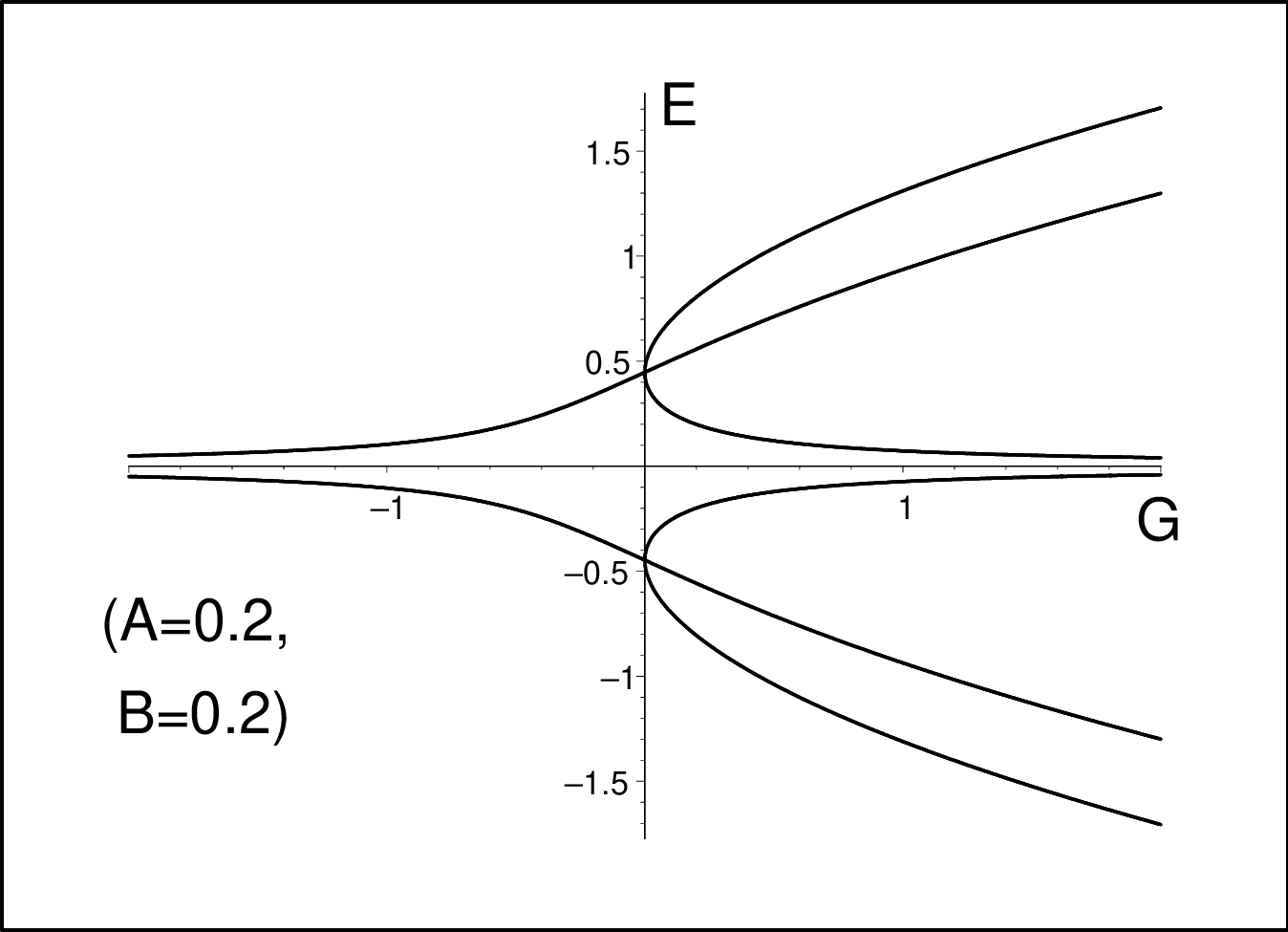,angle=270,width=0.55\textwidth}
\end{center}                         
\vspace{-2mm}\caption{The $N=6$ spectrum near the $B=0$ horizon of
the first kind.
 \label{fia1vc}}
\end{figure}
%

\section{Conclusions\label{summary}}

Let us summarize that in applications of quantum theory the
specification of the physical domain ${\cal D}$ of parameters may be
understood in two ways. A parameter may vary in Hamiltonian
$H=H(\lambda)$ itself ({\em plus}, naturally, in the related
physical Hilbert-space metric) {\em or} solely in the physical
Hilbert space metric (remember that the choice of the
Hamiltonian-Hermitizing metric $\Theta=\Theta(\lambda,\kappa)$ is
not unique in general \cite{Geyer}).

In the former case people often assume that the pass of the quantum
system in question through the EP boundary $\partial {\cal D}$ leads
to the complexification of some energies so that the
unitarity of the evolution is inadvertently lost. In our present
paper we considered the second possibility in which the pass through
the EP boundary {\em does not destroy} the reality of the energies.

We imagined that in such a case one must ask the following natural
question: ``Does this imply that the unitarity of the evolution is
preserved?'' A nontriviality of this question lies in the
fact that after the pass through EP, the very definition of the norm
of the wave functions may change.

By means of a constructive analysis of a few solvable models we
managed to demonstrate that in some cases when
boundary merely separates two disjoint
physical subdomains  ${\cal D}_\pm$
the change of the
definition of the norm of the wave functions is unavoidable.
The value of the norm of a given
wave function performs, in general, a jump when crossing such an EP
horizon $\partial {\cal D}$ of the second kind. In
such a dynamical scenario it is necessary to speak
about a phase transition of the second kind.

We described the mechanism in more detail.
Keeping in mind the popularity of the phase transition of the first
kind (during which the change of the metric is accompanied by the
necessary change of the effective Hamiltonian) we emphasized the
contrast. We introduced the concept of the phase transition
of the second kind during which the change of the metric is {\em
not} accompanied by {\em any} change of the effective Hamiltonian.
Subsequently we emphasized that the change of the physics is
subtler, mediated merely by the change of the physical Hilbert
space, with all of its well known consequences for non-Hamiltonian
observables.

In the related literature one often finds the phase transition of
the first kind interpreted as a symptom of a
spontaneous breakdown of the ${\cal PT}$ symmetry of the system
\cite{Carl}. Via our illustrative examples we demonstrated that the
spontaneous breakdown of the ${\cal PT}$ symmetry
is not necessary for the existence of quantum phase transition.
A ``no-complexification'' dynamical scenario may exist
during which the phase transition does not require any {\em lasting}
loss of ${\cal PT}$ symmetry.

The possibility seems anomalous because after
the system passes through the singularity $\lambda^{(EP)}$, the
Hamiltonian survives without any changes. The most amazing
consequence of the phase transition of
the second kind may be seen in the loss of the observability status of
multiple  operators of observables. The crypto-Hermiticity of
many of them will only hold before or after
the transition.
In any case, the
occurrence of the phase transition of {\em both} kinds will change
the physics thoroughly.

\newpage

\subsection*{Acknowledgements}

D.B. was partially supported by grant of RFBR, grant of President of
Russia for young scientists-doctors of sciences (MD-183.2014.1) and
Dynasty fellowship for young Russian mathematicians. M.Z. was
supported by RV O61389005.


\newpage



\begin{thebibliography}{00}

\bibitem{Carl}
C. M. Bender, Rep. Prog. Phys. 70 (2007) 947.

\bibitem{Ahmed}
Z. Ahmed, D. Ghosh, J. A. Nathan and G. Parkar,
Phys. Lett. A 379 (2015) 2424;

M. Znojil,
arXiv:1303.4876
(unpublished).


\bibitem{Borisov}
D. I. Borisov,
Acta
Polytech. 54 (2014) 93
(arXiv:1401.6316).

\bibitem{ptho}
%
%
M. Znojil,
Phys. Lett. A 259 (1999) 220.

\bibitem{Kato}
T. Kato, Perturbation theory for linear operators, Springer-Verlag,
Berlin, 1966.

\bibitem{catast}
M. Znojil,
J. Phys. A: Math. Theor. 45 (2012) 444036;

Y. N. Joglekar, C. Thompson, D. D. Scott and G. Vemuri,
Eur. Phys. J. Appl. Phys. 63
(2013) 30001;

D. E. Pelinovsky, P. G. Keverekidis and D. J. Frantzeskakis,
Eur. Phys. Lett. 101 (2013) 11002;

C. H. Liang, D. D. Scott and Y. N. Joglekar,
Phys. Rev. A 89 (2014) 030102(R);

D. I. Borisov, F. Ruzicka and M. Znojil,
Int. J. Theor. Phys., in print,
http://dx.doi.org/10.1007/s10773-014-2493-y
(arXiv:1412.6634).

\bibitem{maximal}
M. Znojil,
J. Phys. A: Math. Theor. 40 (2007) 4863;

M. Znojil,
J. Phys. A: Math. Theor. 40 (2007) 13131.


\bibitem{117}
%
A. Mostafazadeh,
J. Phys. A: Math. Gen. 39 (2006) 10171;
%

C. F. de Morison Faria and A, Fring,
Czech. J. Phys. 56 (2006) 899;

V. Jakubsky and J. Smejkal,
Czech. J. Phys. 56 (2006) 985;

A. Ghatak and B. P. Mandal, Comm. Theor. Phys. 59 (2013) 553.


\bibitem{Zelezny}
D. Krej\v{c}i\v{r}\'{\i}k,  P. Siegl and J. \v{Z}elezn\'y,
Complex Anal. Oper. Theory 8 (2014) 255;


D. C. Brody, Consistency of PT-symmetric quantum mechanics,
 arXiv: 1508.02190.

\bibitem{unbali}
D. Krej\v{c}i\v{r}\'{\i}k and P. Siegl,
J. Phys. A: Math. Theor. 43 (2010) 485204;

F. Bagarello and M. Znojil, J. Phys. A: Math. Theor. 44 (2011)
415305;

D. Borisov and D. Krejcirik,
Asympt. Anal. 76 (2012) 49.

\bibitem{SIGMA}
M. Znojil,
SIGMA 5 (2009)  001 (arXiv overlay: 0901.0700);


M. Znojil,
Int. J. Theor. Phys. 52 (2013) 2038.



\bibitem{Geyer}
F. G. Scholtz, H. B. Geyer and F. J. W. Hahne,
Ann. Phys. (NY) 213 (1992)
 74.
%


\bibitem{shendr}
%
M. Znojil and H. B. Geyer,
Fort. d. Physik 61 (2013) 111.
%



\bibitem{ali}
A. Mostafazadeh,
%
Int. J. Geom. Meth. Mod. Phys.  7 (2010) 1191

\bibitem{SIGMAdva}
%
M. Znojil,
SIGMA {4} (2008) 001 (arXiv overlay: 0710.4432).
%


\bibitem{MZbook}
M. Znojil, in ``Non-Selfadjoint Operators in Quantum Physics:
Mathematical Aspects'', F. Bagarello et al, Eds,  Wiley, Hoboken,
2015, pp. 7 - 58.
%


\bibitem{David}
D. Krej\v{c}i\v{r}\'{\i}k, H. B\'{\i}la and M. Znojil,
J. Phys. A 39 (2006) 10143;

D. Krej\v{c}i\v{r}\'{\i}k,
J. Phys. A: Math. Theor. 41 (2008), 244012;
%


C. M. Bender, K. Besseghir, H. F. Jones, and X. Yin, J. Phys. A:
Math. Theor. 42 (2009), 355301.

\bibitem{Wu}
M. Znojil and J. Wu,
Int. J. Theor. Phys. 52 (2013) 2152.


\bibitem{Levai}
G. Levai and M. Znojil,
Mod. Phys. Lett. A 16 (2001) 1973;

A. Sinha and P. Roy, J. Phys. A: Math. Gen. 39 (2006) L377;

P. Dorey, C. Dunning, A. Lishman and R. Tateo,
J. Phys. A: Math. Theor. 42 (2009) 465302;

G. Levai, J. Phys. A: Math. Theor. 45 (2012) 444020.

\bibitem{horizon}
M. Znojil,
J. Phys. A: Math. Theor. 41 (2008) 244027.

\end{thebibliography}
\end{document}